\documentclass{article}
\usepackage{spconf,amsmath,graphicx,hyperref}
\usepackage{booktabs}   
\usepackage{multirow}   

\title{Text2Move: Text-to-moving sound generation via trajectory prediction and temporal alignment}
\twoauthors
 {Yunyi Liu\sthanks{Work was done while interning at and funded by Dolby Laboratories.}}
	{University of Sydney}
 {Shaofan Yang, Kai Li, Xu Li}
	{Dolby Laboratories}
\begin{document}
%
\maketitle
\begin{abstract}

Human auditory perception is shaped by moving sound sources in 3D space, yet prior work in generative sound modelling has largely been restricted to mono signals or static spatial audio. In this work, we introduce a framework for generating moving sounds given text prompts in a controllable fashion. To enable training, we construct a synthetic dataset that records moving sounds in binaural format, their spatial trajectories, and text captions about the sound event and spatial motion. Using this dataset, we train a text-to-trajectory prediction model that outputs the three-dimensional trajectory of a moving sound source given text prompts. To generate spatial audio, we first fine-tune a pre-trained text-to-audio generative model to output temporally aligned mono sound with the trajectory. The spatial audio is then simulated using the predicted temporally-aligned trajectory. Experimental evaluation demonstrates reasonable spatial understanding of the text-to-trajectory model. This approach could be easily integrated into existing text-to-audio generative workflow and extended to moving sound generation in other spatial audio formats.

\end{abstract}


\section{Introduction}
\label{sec:intro}

Spatial audio plays an essential role in our everyday life, from navigating the sound source to augmenting immersive experiences for audio-visual media. In recent years, with the development of generative artificial intelligence and multi-modal learning, modelling spatial audio using another modality such as video~\cite{360} or text~\cite{sun2024both} is emerging to gain more attention. In the domain of text-to-spatial sound modelling, existing works focus on introducing end-to-end architectures generating either the four-channel First-order Ambisonics (FOA) waveforms~\cite{immersediffusion, diff-sage} or binaural audio~\cite{li_tas_2024, sun2024both, pan2025in_the_wild}. However, most of the spatial attributes captured in these works were only static, meaning that the sound source remained still throughout time. In reality, many sound sources are constantly moving while emitting sound waves.


A common approach towards moving sound modelling is via object-based audio, where a mono sound source and its spatial-temporal metadata are used to simulate its movements~\cite{koyama2025past}. Such metadata usually captures essential spatial trajectories of the sound source throughout time, such as azimuth, elevation, distance, reverberation, etc. Parallel to the developments in audio generation, research on trajectory and motion prediction from language or vision inputs has advanced in fields such as robotics~\cite{robot}, autonomous driving~\cite{driving}, and human motion~\cite{motion} synthesis. Models in these domains demonstrate that text~\cite{motion, speaktrajectory} or vision~\cite{morris2008survey, moon2024visiontrap} can provide strong semantic priors for predicting motion patterns in space. Yet, this paradigm has not been fully exploited in the context of audio trajectories, where motion paths can serve as explicit conditioning for spatialization.

In this paper, we introduce a new framework for moving sound generation. Our approach addresses the research gap by decomposing the problem into two components: (i) a text-to-trajectory prediction model, which outputs the spatial-temporal trajectory of an audio object given text prompts, and (ii) a synchronized text-to-audio pipeline, where a fine-tuned generative model produces temporally aligned mono sound that is later spatialized according to the predicted trajectory. To support training and evaluation, we construct a synthetic dataset consisting of (a) binaural audio simulated using HRTF-based rendering, (b) GPT-augmented textual captions describing sound events and their spatial attributes, and (c) ground-truth trajectories parameterized by pre-defined spatial attributes. To the best of our knowledge, this work is the first to explicitly bridge text, trajectories, and audio in a unified framework for spatial audio generation.


\begin{figure*}[h]
  \centering
  \centerline{\includegraphics[width=\textwidth]{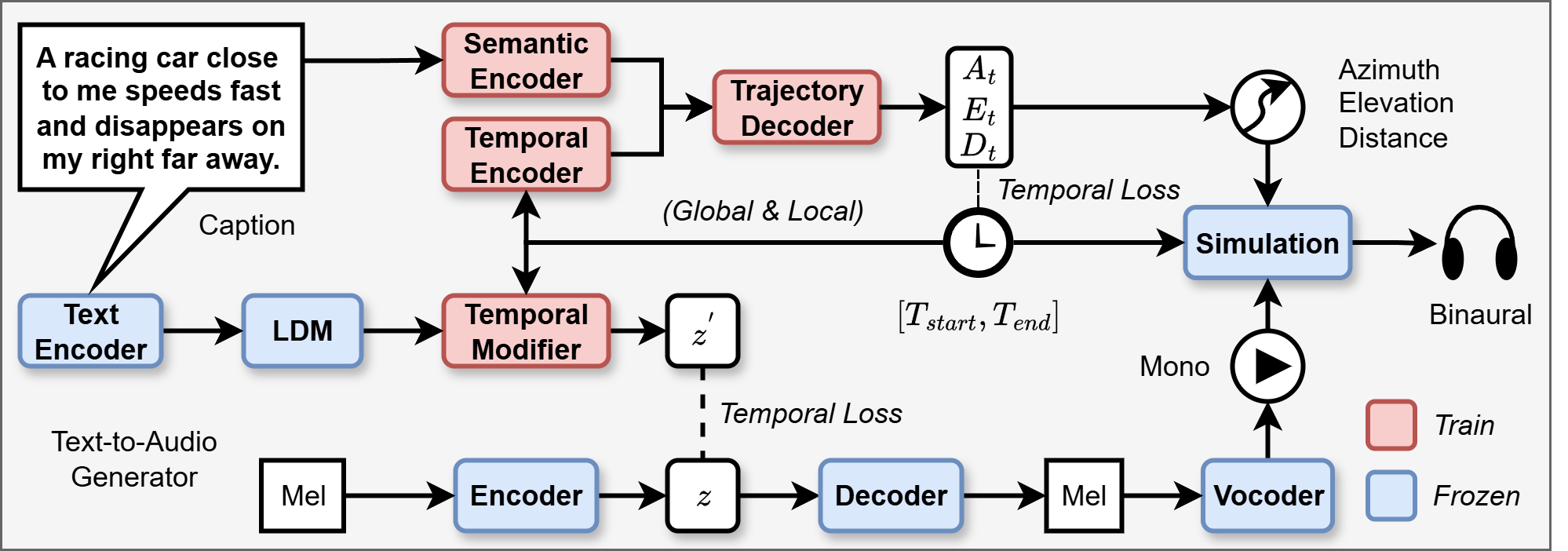}}
  \caption{Proposed model architecture}
  \label{fig: model}
\end{figure*}

\section{Methodology}
\label{sec:method}



\subsection{Text-to-trajectory prediction}

To accurately model a moving sound source, we need to understand its pointwise spatial locations. Our proposed framework learns to predict time-aligned spatial trajectories of a moving sound source from textual descriptions. As shown in Figure\ref{fig: model}, our text-to-trajectory prediction model integrates three key components: a text semantic encoder, a temporal encoder, and a trajectory decoder.

\textbf{Text encoder.} We employ a DistilBERT backbone~\cite{sanh2020distilbertdistilledversionbert} (hidden size 768) as our primary text encoder. Input captions are tokenized and passed through the transformer to produce contextualized embeddings. We then apply a learnable attention pooling vector and project it to a 512-dimensional semantic vector, which forms as a shared latent space (via Linear\,$\rightarrow$\,GELU\,$\rightarrow$\,LayerNorm). This design enables the model to capture semantic attributes from captions that describe the relative spatial movements of the sound source.


\textbf{Temporal encoder.} Temporal information is injected at both global and local levels. At the global level, the encoder takes in the start and end timestamps $(t_0,t_1)$ of an event, which are expanded through Fourier feature mappings with $F{=}8$ log-spaced frequencies to capture multi-scale periodic patterns. These are passed through a lightweight two-layer MLP (32\,$\rightarrow$\,256\,$\rightarrow$\,512 with GELU and LayerNorm) to yield a 512-dimensional temporal embedding. At the local level, each step along the trajectory is assigned a normalized index $\tau\in[0,1]$, which is similarly encoded with Fourier features and projected into a compact 64-dimensional per-step embedding. This provides the model with explicit knowledge of event duration and temporal alignment without requiring frame-level supervision.



\textbf{Trajectory decoder.} The decoder is a transformer model that operates over the temporal dimension using a 4-layer transformer encoder, each configured with 8 attention heads, hidden size 512, feed-forward expansion factor 4, and dropout 0.1. At each timestep, the input is the concatenation of the 512-dimensional text embedding and the temporal embedding, as well as the 64-dimensional per-step features, which are then projected into a shared hidden space before being processed by self-attention layers. A lightweight regression head (two linear layers with GELU and LayerNorm) maps the hidden states to per-step outputs: azimuth, elevation, and distance. To ensure physically meaningful predictions, azimuth and elevation are constrained using $\tanh$ activations scaled by fixed multipliers ($\pm180^\circ$ and $\pm90^\circ$), while distance is constrained with a softplus transformation to enforce positivity.

\textbf{Loss function.} Our training optimizes two objectives. The first is a trajectory loss that computes masked mean absolute error over the predicted and ground-truth curves, using circular L1 error for angular dimensions and standard L1 for distance. At each time step $t$, the model predicts azimuth $\hat a_t$, elevation $\hat e_t$, and distance $\hat d_t$, while the ground truth is $a_t,e_t,d_t$. We introduce a validity mask $m_t \in \{0,1\}$ to restrict the computation to valid steps, and define the circular error as $\Delta^\circ(x,y)=\min\big(|x-y|,\,360-|x-y|\big)$. The three coordinates are balanced by weights $w_{\text{az}},w_{\text{el}},w_{\text{ds}}>0$. The trajectory loss then penalizes the per-step errors over the full (padded) trajectory:
\begin{equation}
    \mathcal{L}_{\text{traj}}
=\frac{\sum_t m_t\Big(
w_{\text{az}}\Delta^\circ(\hat a_t,a_t)+
w_{\text{el}}\Delta^\circ(\hat e_t,e_t)+
w_{\text{ds}}|\hat d_t-d_t|
\Big)}{\sum_t m_t}.
\end{equation}

The second is a temporal loss that encourages the first and last predicted positions to match the annotated start and end points to ensure accurate temporal alignment:
\begin{align}
\mathcal{L}_{\text{time}}
&= w_{\text{az}}\!\left[\Delta^\circ(\hat a_s,a_s)+\Delta^\circ(\hat a_e,a_e)\right] \nonumber \\
&\quad + w_{\text{el}}\!\left[\Delta^\circ(\hat e_s,e_s)+\Delta^\circ(\hat e_e,e_e)\right] \nonumber \\
&\quad + w_{\text{ds}}\!\left[|\hat d_s-d_s|+|\hat d_e-d_e|\right].
\end{align}

The total loss is a weighted sum of both terms:
\begin{equation}
    \mathcal{L}_{\text{total}}=\mathcal{L}_{\text{traj}}+\lambda_{\text{time}}\,\mathcal{L}_{\text{time}},
\end{equation}

where $\lambda_{\text{time}}\!\ge\!0$ controls the strength of the endpoint constraint.

\subsection{Naive spatial metadata prediction}

In addition to the trajectory-based framework, we also established a simpler baseline model that directly predicts six spatial parameters from text, namely the start and endpoints of  azimuth, elevation, and distance. In this formulation, temporal encoder is discarded and the task is simplified to regressing the fixed endpoints (start and end) of a spatial trajectory. We use the same semantic encoder as above followed by a projection layer and a lightweight transformer encoder that models contextual dependencies within the text representation. The resulting embedding is passed to a regression head that outputs six normalized parameters. During training, predictions are optimized using a mean squared error (MSE) loss against the ground truth metadata. We use this design to understand the performance between naively regressing endpoints and our proposed trajectory prediction model.

\subsection{Temporal Alignment with T2A}

While text-to-audio (T2A)~\cite{audioldm2} diffusion models have demonstrated strong generative capabilities, they often lack precise control over when an event occurs within the generated sound. To address this, we introduce a lightweight temporal alignment mechanism that adjusts latent representations from the diffusion model to respect user-provided timing constraints.

In our setup, a pre-trained latent diffusion model (Make-an-Audio 2)~\cite{huang_make--audio_2023} serves as the text-to-audio backbone. Given a text caption and its structured description, the model generates a latent audio representation that can be used to synthesize spectrograms with a transformer-based Variational Autoencoder model. We augment this with a Temporal Modifier, a lightweight trainable module that refines the latent sequence $z_{\text{ldm}}$ based on start and end timestamps $(t_0, t_1)$. The Modifier combines convolutional layers for local temporal smoothing with an MLP that encodes the timestamp pair into a global temporal bias. We then use a binary mask to ensure that only regions outside the target temporal window are modified, while leaving the inside region unchanged. The training objective is a masked mean squared error (MSE), applied only to the outside region:
\[
\mathcal{L}_{\text{temp}}
= \frac{\sum_{t} M_{\text{out},t}\,\big(z_{\text{adj},t} - z_{\text{vae},t}\big)^2}
       {\sum_{t} M_{\text{out},t}} ,
\]
where $z_{\text{vae}}$ is the ground-truth latent representation produced from VAE, $M_{\text{out},t}=1$ if $t$ is outside $[t_0,t_1]$ and $0$ otherwise. This formulation encourages the Modifier to match the ground-truth latents in all regions where no sound event is expected, while preserving the original latents inside the event boundaries. As a result, this model learns to output arbitrary-length sound effects that are temporally matched with our trajectory prediction model by explicitly conditioning the Temporal Modifier on the start and end timestamps.



\subsection{Dataset}

\begin{table}[ht]
\centering
\small
\begin{tabular}{lll}
\toprule
\textbf{Attribute} & \textbf{Caption} & \textbf{Value Range} \\
\midrule
\multirow{7}{*}{Azimuth ($^\circ$)} 
 & left        & $(-100, -80)$ \\
 & front left  & $(-35, -55)$ \\
 & front (OM)      & $(-10, 10)$ \\
 & front right & $(35, 55)$ \\
 & right       & $(80, 100)$ \\
 & right back  & $(125, 145)$ \\
 & back        & $(-170, -180), (170, 180)$ \\
 & left back   & $(-125, -145)$ \\
\midrule
\multirow{3}{*}{Elevation ($^\circ$)} 
 & Up          & $(70, 90)$ \\
 & Middle (OM) & $(-10, 10)$ \\
 & Down        & $(-70, -90)$ \\
\midrule
\multirow{4}{*}{Distance (m)} 
 & Very close  & $(0.3, 0.6)$ \\
 & Close       & $(0.5, 1)$ \\
 & Moderate(OM)& $(1, 3)$ \\
 & Far         & $(3, 10)$ \\
\bottomrule
\end{tabular}
\caption{Dataset mapping of spatial attributes to human-readable captions and value ranges.}
\label{tab:spatial_ranges}
\end{table}

\begin{figure}[ht]
  \centering
  \centerline{\includegraphics[width=\linewidth]{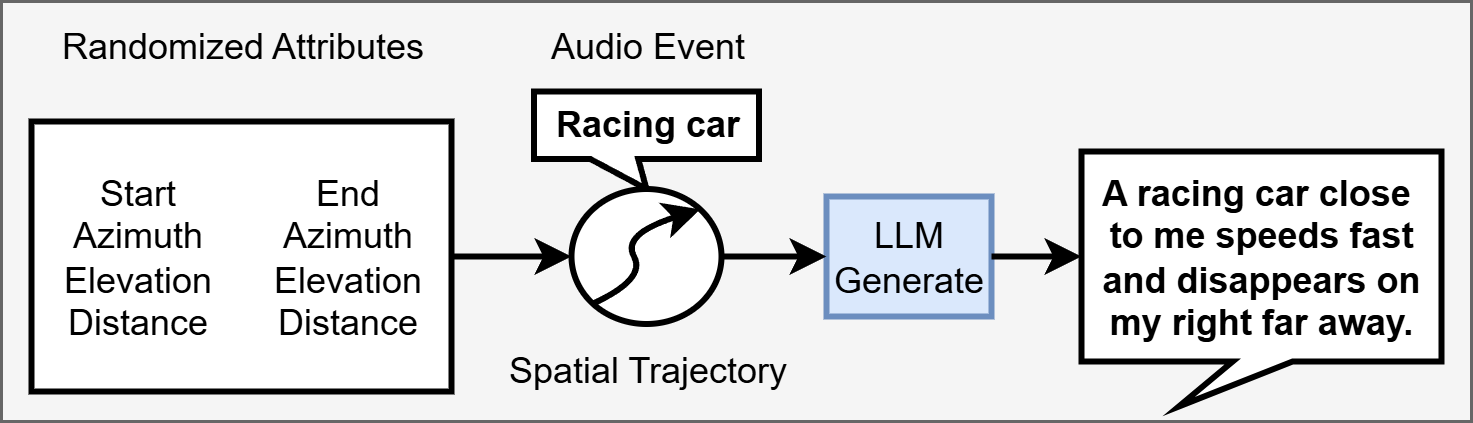}}
  \caption{Dataset curation process}
  \label{fig: dataset}
\end{figure}

Due to the lack of large-scale datasets containing spatial audio with explicit trajectories, we curated a synthetic dataset tailored for text-to-trajectory training. Our work builds upon the AudioTime dataset~\cite{audiotime}, which consists of 5,000 mono audio clips with precise timestamp annotations. Since AudioTime includes clips with multiple overlapping events, we separate the data to contain only single-source events, yielding 7,685 clean clips. To simulate spatial trajectory, we randomly assigned start and end positions for each event using azimuth, elevation, and distance categories derived from human-perceptual ranges (see Fig.~\ref{tab:spatial_ranges}), similar to ~\cite{sun2024both}. This randomization process was repeated ten times per file, resulting in a corpus of 76,850 spatialized samples (213 hours, 90\%/10\% train/test split). Each trajectory corresponds to a source moving linearly from its assigned start to end position at a constant speed with a sampling rate of 20Hz. We then simulate the binaural moving audio by convolving the audio source with the HRIR in each frame using an HRTF library~\cite{armstrong2018perceptual}. To provide a natural caption about the spatial motion, we used GPT-4 to write human-readable text descriptions according to the audio event and spatial attributes. We encouraged lexical diversity by requiring synonyms for audio events, directions, and distance descriptions. For sanity reasons, we ramdonly (50\%) omit captions (OM) where common attributes do not require explicit description. This dataset serves as the foundation for training and evaluating our proposed text-to-moving sound framework.


\subsection{Training}

We trained the text-to-trajectory model with AdamW ($\text{lr}=1\times10^{-5}$, weight decay $1\times10^{-4}$) for $10{,}000$ epochs using a batch size of $256$, mixed-precision training, and gradient clipping at $1.0$. The objective combined masked mean absolute error on azimuth, elevation, and distance with an auxiliary start–end alignment loss. For the temporal alignment model, we froze the latent diffusion backbone and optimized only the temporal adjuster with AdamW ($\text{lr}=1\times10^{-4}$) for $10$ epochs, using a masked reconstruction loss applied outside the annotated temporal window.

\section{Results}
\label{sec:results}

\subsection{Trajectory Prediction}

\begin{table}[t]
\centering
\caption{Prediction results of naive and whole-trajectory model. Metrics include range-aware classification (Accuracy, Macro-F1) and regression (MAE and range-aware MAE).}
\label{tab:compare_models}
\resizebox{\linewidth}{!}{%
\begin{tabular}{lcccc}
\toprule
\textbf{Attribute} & \textbf{Accuracy} & \textbf{Macro-F1} & \textbf{MAE} & \textbf{RA-MAE} \\
\midrule
\multicolumn{5}{l}{\textbf{Naive Model (End Point)}} \\
\midrule
Azimuth   & 98.2\% & 98.5\% & 5.789$^\circ$ & 2.339$^\circ$ \\
Elevation & 98.0\% & 98.1\% & 6.431$^\circ$ & 1.352$^\circ$ \\
Distance  & 87.5\% & 87.1\% & 0.166 & 0.013 \\
\midrule
\multicolumn{5}{l}{\textbf{Whole-trajectory prediction}} \\
\midrule
Azimuth   & 75.9\% & 75.1\% & 18.53$^\circ$ & 15.52$^\circ$ \\
Elevation & 61.2\% & 65.9\% & 28.75$^\circ$ & 21.44$^\circ$ \\
Distance  & 66.7\% & 52.1\% & 1.601 & 0.365 \\
\bottomrule
\end{tabular}
}
\end{table}

In Table~\ref{tab:compare_models}, we show the evaluation results using the test dataset. The naive model predicts near-perfect metadata in most metrics, which demonstrates the capability of our text encoder in understanding spatial semantics. Our proposed trajectory prediction model shows that it is possible to generate temporally aligned spatial motion directly from text. Although the accuracies are lower than the simplified naive adaptation, the MAE suggests a reasonable deviation range (18.53$^\circ$ for Azimuth and 28.75$^\circ$ for elevation). For distance prediction, we observed larger deviations, but we hypothesize this is because the spatial coordinates vary in ranges in our dataset. Therefore, we report the Range-Aware MAE to account for the drifts in prediction: 
\begin{equation}
\text{RA-MAE} = \frac{1}{N} \sum_{i=1}^{N} 
\min_{r \in \mathcal{R}} \; \Big| \hat{y}_i - \Pi_r(y_i) \Big|,
\end{equation}
where $\hat{y}_i$ is the predicted value, $y_i$ is the ground-truth, and $\mathcal{R}$ denotes the set of predefined valid ranges for the attributes~\ref{tab:spatial_ranges}. The operator $\Pi_r(y_i)$ projects $y_i$ into its corresponding range $r \in \mathcal{R}$, so the error is measured relative to the nearest boundary within that range. We found the predicted trajectories in general fall within the boundaries of the pre-defined ranges, although it slightly drifts slightly in an acceptable scope.

\subsection{Temporal Alignment}

\begin{table}[t]
\centering
\caption{Temporal alignment performance of our approaches. \emph{Start} and \emph{End MAE} measure the average timing error (in seconds) at the predicted onset and offset boundaries. \emph{OLR} (Overlap Ratio) quantifies the overall temporal overlap between predicted and ground-truth intervals.}
\label{tab:temporal_alignment}
\begin{tabular}{lccc}
\toprule
\textbf{Method} & \textbf{Start MAE} & \textbf{End MAE} & \textbf{OLR} \\
\midrule
Trajectory predictor & 0.0086 & 0.0012 & 0.8596 \\
Temporal modifier & 0.0018 & 0.0024 & 0.9370 \\
\bottomrule
\end{tabular}
\end{table}

As shown in Table~\ref{tab:compare_models}, both our trajectory predictor and temporal modifier models are able to output highly accurate temporal alignments (below 10\,ms absolute error between the ground-truth and predicted start/end timestamps). The OLR (Overlap Ratio) measures the overlap between the predicted and ground-truth active windows: 
\begin{equation}
\mathrm{OLR}_{\text{mask}}
= \frac{\sum_{t=1}^{T} m_{\text{pred}}(t)\, m_{\text{gt}}(t)}
       {\sum_{t=1}^{T} \max\{m_{\text{pred}}(t),\, m_{\text{gt}}(t)\}} \;\in [0,1].
\end{equation}
where $m_{\text{pred}}(t)$ and $m_{\text{gt}}(t)\in\{0,1\}$ are the predicted and ground-truth activity masks(1 = active). An OLR of $0.86$ from the trajectory predictor indicates a strong temporal alignment, although in some regions it deviates slightly. The temporal modifier on the other hand, obtains an even higher OLR of $0.94$, hypothetically due to the introduction of binary masks. These results demonstrate that both approaches are able to deliver reliable and accurate temporal control in general.

\section{Discussion}
\label{sec:discussion}

This research presents text2move, a hybrid approach towards moving sound generation guided by text prompts. Different from end-to-end methods, our approach disentangles the sound generation process into a few simpler steps: text-to-trajectory prediction, mono sound generation and synchronization, as well as object-based spatialization. This framework takes advantage of pre-trained text-to-mono-audio generation models, and provides high-level controllability of modifying the semantics, dynamics and spatial characteristics of moving sounds. By temporally aligning the predicted trajecotry and generative audio, our approach can deliver accurate spatial audio movements demonstrated from our website~\url{https://reinliu.github.io/text2move/}.

\clearpage






\vfill\pagebreak




\bibliographystyle{IEEEbib}
\bibliography{strings,refs, references}

\end{document}